\documentstyle[preprint,tighten,aps,epsfig,colordvi,psfig]{revtex}

\def\lsim{\mathrel{\rlap{\lower4pt\hbox{\hskip1pt$\sim$}}
    \raise1pt\hbox{$<$}}}                
\def\gsim{\mathrel{\rlap{\lower4pt\hbox{\hskip1pt$\sim$}}
    \raise1pt\hbox{$>$}}}                

\begin{document}
\setlength{\textheight}{25cm}
\textheight=23cm
\draft
\title{QCD with overlap fermions: Running coupling \\ and the 3-loop $ \mathbf \beta$-function}
\author{M. Constantinou, H. Panagopoulos\\}
\address{
Department of Physics, University of Cyprus,\\
P.O.Box 20537, Nicosia CY-1678, Cyprus\\
{\it email:}{ phpgmc1@ucy.ac.cy, haris@ucy.ac.cy}}

\maketitle

\begin{abstract}
We calculate the relation between the bare coupling constant $g_0$ and the $\rm \overline{MS}$-renormalized 
coupling $g_{\,\rm \overline{MS}}$, $g_0 = Z_g(g_0,a\mu) g_{\,\rm \overline{MS}}$, to 2 loops in perturbation theory.
We employ the standard Wilson action for gluons and the overlap action for fermions.
For convenience, we have worked with the background field technique, which only requires evaluation of 
2-point Green's function for the problem at hand. 
Our results depend explicitly on the number of fermion flavors ($N_f$) and colors ($N$). 
Since the dependence of $Z_g$ on the overlap parameter $\rho$ 
cannot be extracted analytically, we tabulate our results for different values in the allowed range 
of $\rho$ ($0<\rho<2$), focusing on values which are being used most frequently in simulations. 
Knowledge of $Z_g$ allows us to derive the 3-loop coefficient of the bare $\beta$-function ($\beta_L(g_0)$) which,
unlike the 1- and 2-loop coefficients, is regularization-dependent.
The nontrivial dependence of $Z_g$ and of $\beta_L(g_0)$ on $\rho$ is plotted for various choices of $N$, $N_f$.



\end{abstract}
\newpage

\section{INTRODUCTION}

In later years, use of non-ultralocal actions which preserve chiral symmetry on the lattice has become more viable.
The two actions which are being used most frequently are overlap fermions~\cite{NN,Neuberger-98,Neuberger-01} 
based on the Wilson fermion action and domain-wall fermions~\cite{Kaplan,FS}.

Overlap fermions are notoriously difficult to study, both numerically and analytically. 
Many recent promising investigations involving simulations with overlap fermions have appeared; see, e.g., Refs.~\cite{DLS,IKKSSW,BFLW,JLQCD,BGHHLR,BJSSS,DMZACDHLLT}.  
Regarding analytical computations, the only ones performed thus far have been either up to 1 loop, such as Refs.~\cite{GHPRSS,IP,D,Capitani,CG,AFPV,APV},
or vacuum diagrams at higher loops~\cite{SP,AP}.
{\it{The present work is the first one involving non-vacuum diagrams beyond the 1-loop level.}}

We compute the 2-loop renormalization $Z_g$ of the bare lattice coupling constant $g_0$ in the presence of overlap fermions.
We relate $g_0$ to the renormalized coupling constant $g_{\,\rm \overline{MS}}$ as defined in the ${\rm \overline{MS}}$ scheme
at a scale $\bar{\mu}$; at large momenta, these quantities are related as follows
\begin{equation}
\alpha_{\rm \,\overline{MS}}(\bar{\mu}) = \alpha_0 + d_1(\bar{\mu} a)\alpha_0^2 + d_2(\bar{\mu} a)\alpha_0^3 + ... \,,
\end{equation}
($\alpha_0=g_0^2/4\pi, \, \alpha_{\rm \,\overline{MS}}=g_{\rm \,\overline{MS}}^2/4\pi, \, a:$ lattice spacing).
The 1-loop coefficient $d_1(\bar{\mu} a)$ has been known for a long time; several evaluations of $d_2(\bar{\mu} a)$ 
have also appeared in the past $\sim$10 years, either in the absence of fermions \cite{HHnew,LW}, or using the Wilson 
\cite{CFPVnew} or clover \cite{BPnew,BPPnew} fermionic actions. Knowledge of $d_2(\bar{\mu} a)$, along with the 3-loop
$\rm \overline{MS}$-renormalized $\beta$-function \cite{TVZ} allows us to derive the 3-loop bare lattice $\beta$-function,
which dictates the dependence of lattice spacing on $g_0$. 
In particular, it provides a correction to the standard 2-loop asymptotic scaling formula defining $\Lambda_L$.
Ongoing efforts to estimate the running coupling from the lattice \cite{MTDFGLNS,Sommer,ALPHA,KKPZ} 
have relied on a mixture of perturbative and non-perturbative investigations. As a particular example, relating 
$\alpha_{\rm \,\overline{MS}}$ to $\alpha_{\rm SF}$ (SF: Schr\"odinger Functional scheme, as advocated 
by the ALPHA Collaboration), entails an intermediate passage through the bare coupling and the conversion
from $\alpha_{\rm \,\overline{MS}}$ to $\alpha_0$ is carried out perturbatively.

The paper is organized as follows:
Some theoretical background and the methodology of the necessary perturbative calculations are given in Section II. 
Section III regards the overlap action where the derivation of the vertices with up to 4 gluons is explained.
 We also provide the expressions for these vertices (the 4-gluon vertex is given in Appendix A). 
Details on our computation, numerical results and plots of $Z_g$ coefficients and the $\beta$-function can be found in Section IV. 
Finally, in Section V we give the summary and our conclusions. 

The present study, being the first of its kind in calculating 2-loop diagrams with overlap vertices and external momentum dependence,
had a number of obstacles to overcome. One first complication is the size of the algebraic form of the Feynman vertices;
as an example, the vertex with 4 gluons and a fermion-antifermion pair contains $\sim$724,000 terms when expanded.
Upon contraction these vertices lead to huge expressions (many millions of terms); this places severe requirements 
both on the efficiency of the computer algorithms which we must design to manipulate such expressions automatically
and on the necessary computer RAM.
Numerical integration of Feynman diagrams over loop momentum variables is performed on a range of lattices, with finite size $L$, 
and subsequent extrapolation to $L\rightarrow \infty$. As it turned out, larger $L$ were required for an accurate
extrapolation in the present case, compared to ultra-local actions.
In addition, since the results depend nontrivially on the parameter $\rho$ of the overlap action, numerical evaluation must be
performed for a sufficiently wide set of values of $\rho$, with an almost proportionate increase in CPU time.
Extreme values of $\rho$ ($\rho\gsim 0,\, \rho\lsim 2$) show unstable numerical behaviour, which is attributable to the spurious poles of the fermion propagator
at these choices; this forces us to even larger $L$. A consequence of all complications noted above is an extended use of CPU time:
Our numerical integration codes, which ran on a 32-node cluster of dual CPU Pentium IV processors, required a total of $\sim$50 years of CPU time.


\section{THEORETICAL BACKGROUND}

The definition and value of the renormalized coupling constant $g$ depends on the renormalization scheme (parameterized by a scale $\mu$), and this 
dependence is given by the renormalized $\beta$-function
\begin{equation}
\displaystyle \beta (g) \equiv \mu \frac{dg}{d\mu}
\end{equation}
We will adopt the $\rm \overline{MS}$ renormalization scheme throughout this work, and we will denote the renormalized coupling constant $g_{\,\rm \overline{MS}}$ simply by $g$.
For the lattice regularization a bare $\beta$-function is defined as
\begin{equation}
\displaystyle \qquad\qquad  \beta_L(g_0)= -a{dg_0\over da} \Bigg|_{g,\,\bar{\mu}}
\end{equation}
where $\bar{\mu}$ the renormalization scale and $g$ $(g_0)$ the renormalized (bare) coupling constant.
It is well known that in the asymptotic limit for QCD ($g_0\rightarrow 0$), one can write the expansion of the $\beta$-function
in powers of $g_0$, that is
\begin{equation}
\beta_L(g_0) =-b_0 \,g^3_0 -b_1 \,g_0^5 - b_2^{L}\,g_0^7 - ... 
\end{equation}
The coefficients $b_0, b_1$ are universal constants (regularization independent) given by
\begin{eqnarray}
&&b_0 = {1\over (4\pi)^2} 
\left({11\over 3}N-{2\over 3}N_f\right)\nonumber \\
\nonumber \\
&&b_1= {1\over (4\pi)^4} \left[{34\over 3}N^2 - N_f \left(
{13\over 3}N- {1\over N}\right)\right]
\label{univ_coeff}
\end{eqnarray}
where $N (N_f)$ is the number of colors (flavors).\\
On the contrary, $b_i^{L}$ ($i \ge 2 $) depends on the regulator; it must be determined perturbatively.
In the present work we calculate the coefficient $b_2^{L}$ using the $\rm overlap$ fermionic action and Wilson gluons.

The perturbative expansion of the renormalized $\beta$-function is
\begin{equation}
\beta(g) = \phantom{*} \bar{\mu}{dg\over d\bar{\mu}} \Bigg|_{a,g_0}\,\,  
=\,\,  -b_0 \,g^3 -b_1 \,g^5 - b_2\,g^7 + ... 
\end{equation}
$\displaystyle \beta_L(g_0)$ and $\displaystyle \beta(g)$ can be related using the renormalization function 
$Z_g$, defined through $g_0 = Z_g(g_0,a\bar{\mu}) g\,$, that is
\footnote{$Z_g$ could be denoted: $Z_g^{L,{\rm \,\overline{MS}}}$ to indicate its dependence on 
the regulator ($L$: lattice) and on the renormalization scheme ($\rm \overline{MS}$).}
\begin{equation}
\beta^{L}(g_0) = \left( 1 - g_0^2 \,\,{\partial \ln Z_g^2 \over \partial g_0^2}  \right)^{-1}Z_g \,\,\beta(g_0Z_g^{-1})
\label{beta1}
\end{equation}
Computing $Z_g^2$ to 2 loops
\begin{eqnarray}
&&Z_g^2(g_0,a\bar{\mu})  = 1 + g_0^2\,(2b_0 \ln(a\bar{\mu})+ l_0) +g_0^4\,(2b_1 \ln(a\bar{\mu})+ l_1) + O(g_0^6) 
\label{Zg}
\end{eqnarray}
and inserting it in Eq. (\ref{beta1}), allows us to extract the 3-loop coefficient $b_2^L$. 
The quantities $b_0, b_1,b_2$ and $l_0$ have been known in the literature 
for quite some time~\cite{TVZ,APV}; $b_0$ and $b_1$ are the same as those of the bare $\beta$-function, 
Eq. (\ref{univ_coeff}), and $b_2$ in the $\rm\overline{MS}$ scheme is
\begin{equation}
b_2 = {1\over (4\pi)^6}
\left[ {2857\over 54}N^3 + N_f \left( -{1709N^2\over 54} + {187\over 36}
+{1\over 4N^2}\right) + 
N_f^2 \left( {56 N \over 27} - {11\over 18N}\right)\right]
\label{b2MS}
\end{equation}
The constant $l_0$ is related to the ratio of the $\Lambda$ parameters associated with 
the particular lattice regularization and the $\overline{\rm MS}$ renormalization scheme
\begin{equation}
l_{0} = 2b_0\ln \left( \Lambda_L/
\Lambda_{\,\overline{\rm MS}}\right)
\end{equation}
For overlap fermions the exact form of $l_0$ appears in Ref.~\cite{APV}
\begin{equation}
l_0 = {1\over 8 N} - 0.16995599 N
+ N_f\left[ - {5\over 72 \pi^2} - k(\rho)\right]
\label{l0}
\end{equation}
where $k(\rho)$ is the convergent part of the 1-loop fermionic contribution 
(denoted by $k_f(\rho)$ in Ref.~\cite{APV}), presented in Table I.

Eq.~(\ref{beta1}) is valid order by order in perturbation theory and 
expanding it in powers of $g_0^2$ the first nontrivial relation is 
\begin{equation}
b_2^L= b_2 -b_1l_{0}+ b_0 l_{1}
\label{b2lrel}
\end{equation}
Thus, the evaluation of $b_2^{L}$ requires only the determination of the 2-loop quantity $l_{1}$.

A direct outcome of our calculation is the 2-loop corrected asymptotic scaling relation between $a$ and $g_0$
\begin{equation}
a = {1\over\Lambda_L} \exp \left( -{1\over 2b_0g_0^2}\right)
(b_0g_0^2)^{-{b_1/2b_0^2}}
\left[ 1 + q\, g_0^2 + {\cal O}(g_0^4)\right], \qquad
q = {b_1^2-b_0b_2^L\over 2b_0^3}
\label{asympre}
\end{equation}
where all quantities in the correction term $q$, except $b_2^L$, are known.

The most convenient and economical way to proceed with calculating $Z_g(g_0,a\bar{\mu})$ 
is to use the background field technique \cite{Abbott,EM,LW}, 
in which the following relation is valid
\begin{equation}
Z_A(g_0,a\bar{\mu})   Z_g^2(g_0, a\bar{\mu}) = 1
\end{equation}
where $Z_A$, defined as: $A^{\mu}(x) = Z_A(g_0,a\bar{\mu})^{1/2} A_{R}^{\mu}(x)$ 
is the background field renormalization function ($A^{\mu}$ ($A_R^{\mu}$): bare (renormalized) background field). 
In the lattice version of the background field technique, the link variable takes the form
\begin{equation}
U_{\mu}(x)= e^{i a g_0 Q_{\mu}(x)}\cdot e^{i a A_{\mu}(x)}
\label{link}
\end{equation}
($Q_{\mu}(x)$: quantum field, $A_{\mu}(x)$: background field).
In this framework, instead of $Z_g(g_0,a\bar{\mu})$, it suffices to compute $Z_A(g_0,a\bar{\mu})$,
with no need to evaluate any 3-point functions.
For this purpose, we consider the background field one-particle irreducible (1PI) 2-point function, 
both in the continuum (dimensional regularization, $\rm \overline{MS}$ subtraction): $\Gamma^{AA}_R(p)^{ab}_{\mu\nu}$ 
and on the lattice: $\Gamma^{AA}_L(p)^{ab}_{\mu\nu}$. 
In the notation of Ref.~\cite{LW}, these 2-point functions can be expressed in terms of scalar functions $\nu_R(p),\,\nu(p)$
\begin{eqnarray}
&&\Gamma^{AA}_R(p)^{ab}_{\mu\nu} = 
-\delta^{ab}\left( \delta_{\mu\nu}p^2 - p_\mu p_\nu\right)
\left( 1 - \nu_R(p)\right)/g^2, \qquad
\nu_R(p) = g^{2} \nu_R^{(1)}(p)+g^{4} \nu_R^{(2)}(p)+... 
\label{2pt-functionA_R}\\
\nonumber \\
&&\sum_\mu \Gamma^{AA}_L(p)^{ab}_{\mu\mu} =  
-\delta^{ab}3\widehat{p}^2 
\left[ 1 - \nu(p)\right]/g_0^2, \qquad\qquad\qquad\quad
\nu(p) = g_0^{2} \nu^{(1)}(p)+g_0^{4} \nu^{(2)}(p)+... 
\label{2pt-functionA}
\end{eqnarray}
($\hat{p}_\mu=(2/a) \sin(ap_\mu/2)$). 
There follows
\begin{equation}
Z_A = {1 - \nu_R(p,\bar{\mu},g)\over 1 - \nu(p,a,g_0)}
\end{equation}
The gauge parameter $\lambda$ must also be renormalized (up to 1 loop), in order to compare lattice and continuum results
\begin{eqnarray}
\lambda = Z_Q\lambda_0 \, , \qquad Z_Q = 1+ g_0^2 z_Q^{(1)}+...
\end{eqnarray}
($Z_Q$: renormalization function of the quantum field). The coefficient $z_Q^{(1)}$ is obtained from the 
quantum field 1PI 2-point function in the continuum ($\Gamma^{QQ}_R(p)^{ab}_{\mu\nu}$) 
and on the lattice ($\Gamma^{QQ}_L(p)^{ab}_{\mu\nu}$) through
\begin{eqnarray}
&&\Gamma^{QQ}_R(p)^{ab}_{\mu\nu} =
-\delta^{ab}
\left[ \left( \delta_{\mu\nu}p^2 - p_\mu p_\nu\right)
\left( 1 - \omega_R(p)\right) + \lambda p_\mu p_\nu \right], \quad\,
\omega_R(p) = g^{2} \omega_R^{(1)}(p)+{\cal O}(g^{4})  
\label{2pt-functionQ_R}\\
\nonumber \\
&&\sum_\mu \Gamma^{QQ}_L(p)^{ab}_{\mu\mu} =
-\delta^{ab}\widehat{p}^2
\left[ 3\left( 1 - \omega(p)\right) + \lambda_0\right], \qquad\qquad\qquad\quad\,
\omega(p) = g_0^{2}\, \omega^{(1)}(p)+{\cal O}(g_0^{4})
\label{2pt-functionQ}
\end{eqnarray}
\vspace{-.45cm}
\begin{equation}
z_Q^{(1)} = \omega^{(1)}(p,a,g_0)  - \omega_R^{(1)}(p,\bar{\mu},g)
\end{equation}
In terms of the perturbative expansions Eqs.~(\ref{2pt-functionA_R}), (\ref{2pt-functionA}), (\ref{2pt-functionQ_R}), (\ref{2pt-functionQ}),
$Z_g^2$ takes the form
\begin{equation}
Z_g^2 = \Big[ 1 + g_0^2\, (\nu_R^{(1)} - \nu^{(1)}) + g_0^4\, (\nu_R^{(2)} - \nu^{(2)}) + 
\lambda_0\, g_0^4\, (\omega^{(1)} - \omega_R^{(1)})\frac{\partial \nu_R^{(1)}}{\partial \lambda} \Big]_{\lambda=\lambda_0}
\label{Zg_expanded}
\end{equation}
The fermion part of $\omega^{(1)}$ coincides with that of $\nu^{(1)}$. Similarly for the fermion part of $\omega_R^{(1)}$ and $\nu_R^{(1)}$.
Consequently, one may write
\begin{eqnarray}
\omega^{(1)} - \omega_R^{(1)} = [\omega^{(1)} - \omega_R^{(1)}]_{N_f=0} + [\nu^{(1)} - \nu_R^{(1)}] - [\nu^{(1)} - \nu_R^{(1)}]_{N_f=0}
\end{eqnarray}
Since the quantities of interest are gauge invariant, we choose to work in the bare Feynman gauge, $\lambda_0=1$, for convenience.
In order to compute $Z_A$ we need the expressions for $\nu_R^{(1)},\,\nu_R^{(2)},\,z_Q^{(1)},\,\nu^{(1)},\,\nu^{(2)}$. 
The $\overline{\rm MS}$ renormalized functions necessary for this calculation to 2 loops are~\cite{LW,AFP}
\begin{eqnarray}
\nu_R^{(1)}(p,\lambda)=&&
{N\over 16\pi^2} \left[  
-{11\over 3}\ln {p^2\over \bar{\mu}^2} +{205\over 36}
+{3\over 2\lambda} + {1\over 4 \lambda^2}\right] +
{N_f\over 16\pi^2}\left[ {2\over 3}\ln{p^2\over \bar{\mu}^2}
- {10\over 9}\right]
\label{nu1}\\
\nonumber \\
\nonumber \\
\omega_R^{(1)}(p,\lambda)=&&
{N\over 16\pi^2} \left[ \left(-{13\over 6}+{1\over 2\lambda}\right) 
\ln{p^2\over \bar{\mu}^2} +{97\over 36}
+{1\over 2\lambda} + {1\over 4 \lambda^2}\right] +
{N_f\over 16\pi^2}\left[ {2\over 3}\ln{p^2\over \bar{\mu}^2}- 
{10\over 9}\right]
\label{omega1}\\
\nonumber \\
\nonumber \\
\nu_R^{(2)}(p,\lambda=1)=&&
{N^2\over \left(16\pi^2\right)^2} 
\left[ -8\ln{p^2\over \bar{\mu}^2} +{577\over 18}-6\zeta(3)\right]+\nonumber \\
\nonumber \\
&&{N_f\over \left(16\pi^2\right)^2}
\left[ N\left( 3 \ln{p^2\over \bar{\mu}^2} - {401\over 36}\right)+ {1\over N}
\left(-\ln{p^2\over \bar{\mu}^2} + {55\over 12} - 4\zeta(3)\right)\right]
\label{nu2b}
\end{eqnarray}

For the lattice quantities, the gluonic contributions ($N_f = 0$) have been presented in previous works~\cite{LW,AFP} 
(for the Wilson action) 
\begin{eqnarray}
\omega^{(1)}(p,\lambda_0=1) =&& -{5N\over 48\pi^2}\ln{(a^2p^2)} -{1\over 8N}+0.137286278291N 
\label{omega1_1} \\
\nonumber \\
\nu^{(1)}(p,\lambda_0=1) =&& -{11N\over 48\pi^2}\ln{(a^2p^2)} -{1\over 8N} + 0.217098494367N 
\label{nu1_1} \\ 
\nonumber \\
\nu^{(2)}(p,\lambda_0=1) =&& -{N\over 32\pi^4}\ln{(a^2p^2)} +{3\over 128N^2}-0.01654461954+0.0074438722N^2 
\label{nu2_1} 
\end{eqnarray}

The fermionic contributions are associated with the diagrams of Fig. 1 and Fig. 2. 
In the present work, $\nu^{(2)}$ is perturbatively calculated for the first time using overlap fermions and Wilson gluons.
For completeness, we also compute the coefficient $\nu^{(1)}$ and compare it with previous results.
The 1-loop diagrams (Fig. 1) correspond to $\nu^{(1)}$, while the 2-loop diagrams (Fig. 2) lead to $\nu^{(2)}$.
Dashed lines ending in a cross represent the background field, while those inside loops denote the quantum field.
Solid lines correspond to fermion fields and a dot stands for the mass counterterm.
Note that, for overlap fermions, the mass counterterm equals zero, by virtue of the exact chiral symmetry of the overlap action;
consequently, diagrams 19 and 20 both vanish.
Certain 2-loop diagrams have infrared divergences and become convergent only  
when grouped together (6+12, 7+11, 8+18, 9+17).

\newpage
\vskip 1cm
\centerline{\psfig{figure=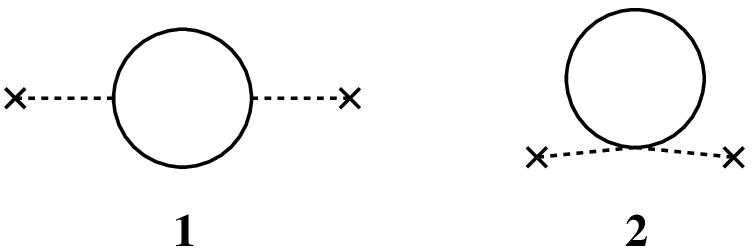,height=2.25truecm}}
\noindent
{\footnotesize {Fig. 1: Fermion contributions to the 1-loop function $\nu^{(1)}$.
Dashed lines ending on a cross represent background gluons.
Solid lines represent fermions.}}\\
\vskip .8cm
\bigskip
\centerline{\psfig{figure=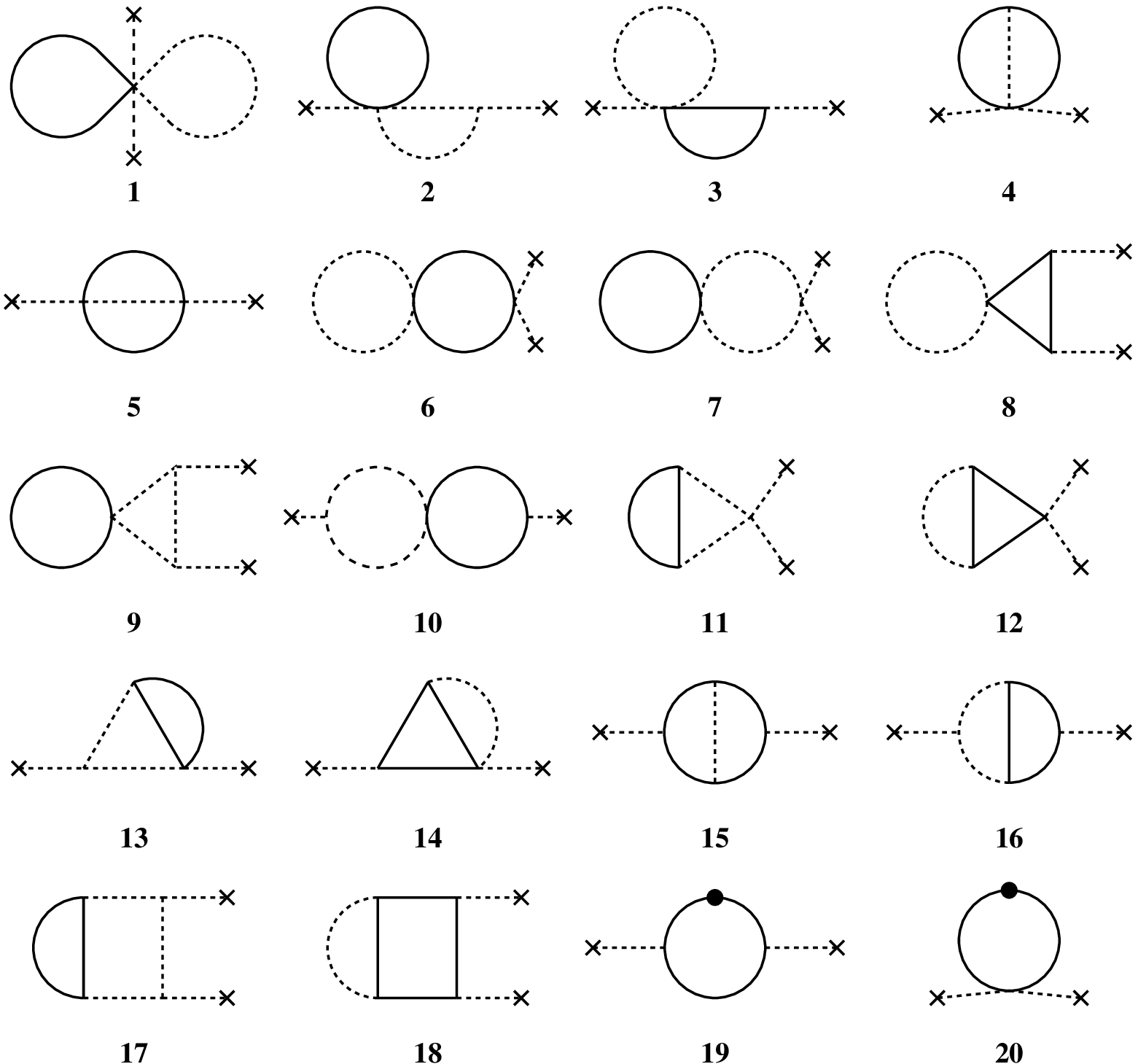,height=14truecm}}
\noindent
{\footnotesize {Fig. 2: Fermion contributions to the 2-loop function $\nu^{(2)}$.
Dashed lines represent gluonic fields;
those ending on a cross stand for background gluons.
Solid lines represent fermions.
The filled circle is a 1-loop
fermion mass counterterm.}}
\vskip .25cm


\section{OVERLAP ACTION}

In recent years, overlap fermions are being used ever more extensively in numerical simulations,
both in the quenched approximation and beyond. 
This fact, along with the desirable properties of the overlap action, was our motivation 
to calculate the $\beta$-function with this type of fermions. 
The important advantage of the overlap action is that it preserves chiral symmetry 
while avoiding fermion doubling. It is also ${\cal O}(a)$ improved. 
The main drawback of this action is that it is necessarily non-ultralocal; as a consequence,
both numerical simulations and perturbative studies are extremely difficult and demanding 
(in terms of human, as well as computer time).

The overlap action is given by~\cite{Neuberger-98}
\begin{equation}
S_{\rm overlap} = a^8 \sum_{n,m} \bar{\Psi}(n) \, D_N (n,m) \, \Psi(m)
\end{equation}
where $D_N (n,m)$ is the overlap-Dirac operator
\begin{eqnarray}
D_N (n,m) &=& \rho \Bigg[\frac{\delta_{n,m}}{a^4}-\left(X\frac{1}{\sqrt{X^\dagger X}}\right)_{nm}\Bigg], \qquad
X = \frac{1}{a^4} \left(D_W -\rho \right)
\end{eqnarray}
and $D_W$ is the Wilson-Dirac operator
\begin{equation}
D_{\rm W} = {1\over 2} \left[ \gamma_\mu \left( \nabla_\mu^*+\nabla_\mu\right)
 - a\nabla_\mu^*\nabla_\mu \right], \qquad
\nabla_\mu\psi(x) = {1\over a} \left[ U(x,\mu) \psi(x + a\hat{\mu})- \psi(x)\right]
\label{derivative}
\end{equation}
The overlap parameter $\rho$ is restricted by the condition $0<\rho <2$ to guarantee the correct pole structure of $D_N$.
The coupling constant is included in the link variables, present in the definition of $X$,
 and one must take the perturbative expansion of $X$ in powers of $g_0$.
This expansion in momentum space takes the form
\begin{equation}
X(p',p)=\underbrace{\chi_0(p)(2\pi)^4 \delta_P(p' - p)}_{tree-level} + 
\underbrace{X_1(p',p)+X_2(p',p)}_{1-loop} +
\underbrace{X_3(p',p)+X_4(p',p)}_{2-loop} + O(g^5_0)
\end{equation}
where $\chi_0$ is the inverse fermion propagator and $X_i$ are the vertices of the Wilson fermion action 
($p$ ($p'$): fermion (antifermion) momentum). 
The construction of all overlap vertices relevant to the present computation 
(see Eqs. (\ref{vertices}), (\ref{V1}), (\ref{V2}), (\ref{V3}), (\ref{V4}) below)
make use of $\chi_0$ and $X_1-X_4$; these quantities can be written in the compact form
\begin{eqnarray}
&&\chi_0(p)=\frac{i}{a}\sum_{\mu}\gamma_{\mu}\sin(ap_{\mu})+\frac{r}{a}\sum_{\mu}\Big(1-\cos(ap_{\mu})\Big)-\frac{\rho}{a} \nonumber \\
&&X_1(p',p)=g_0\int
d^4k\delta(p'-p-k)\sum_{\mu}A_{\mu}(k)V_{1,\mu}\Big(\frac{p'+p}{2}\Big) \nonumber \\
&&X_2(p',p)=\frac{g_0^2}{2}\int
\frac{d^4k_1d^4k_2}{(2\pi)^4}\delta(p'-p-k_1-k_2)\sum_{\mu}A_{\mu}(k_1)
A_{\mu}(k_2)V_{2,\mu}\Big(\frac{p'+p}{2}\Big) \nonumber\\
&&X_3(p',p)=\frac{g_0^3}{3!}\int
\frac{d^4k_1d^4k_2d^4k_3}{(2\pi)^8}\delta(p'-p-\sum_{i=1}^3 k_i)\sum_{\mu}\prod_{i=1}^3 A_{\mu}(k_i)
\Big[-a^2 V_{1,\mu}\Big(\frac{p'+p}{2}\Big) \Big] \nonumber\\
&&X_4(p',p)=\frac{g_0^4}{4!}\int
\frac{d^4k_1d^4k_2d^4k_3d^4k_4}{(2\pi)^{12}}\delta(p'-p-\sum_{i=1}^4 k_i)\sum_{\mu}\prod_{i=1}^4 A_{\mu}(k_i)
\Big[-a^2 V_{2,\mu}\Big(\frac{p'+p}{2}\Big) \Big]
\label{X's}
\end{eqnarray}
where
\begin{eqnarray}
V_{1,\,\mu}(p)=i\,\gamma_{\,\mu}\cos(ap_{\,\mu})+r\,\sin(ap_{\,\mu}), \quad V_{2,\,\mu}(p)=-i\,\gamma_{\,\mu}a\,\sin(ap_{\,\mu})+a\,r\,\cos(ap_{\,\mu})
\end{eqnarray}
$A_{\,\mu}$ represents a gluon field; later on we will have to generalize Eqs. (\ref{X's}) to the case where both
background and quantum gluon field are present, see Eq. (\ref{X_QA}).

At this point we can proceed with the perturbative expansion of $D_N$ in powers of $g_0$. This leads to the propagator of zero mass fermions 
and to gluon-fermion-antifermion vertices (with up to 4 gluons for the needs of the present work).
The much simpler case of vertices with up to 2 gluons (and no background) can be found in Ref.~\cite{KY}.
After laborious analytical manipulations (an essential step is the expansion of $\displaystyle 1 / \sqrt{X^\dagger X}$ 
using complex analysis, which is presented in Appendix A), the overlap-Dirac operator is expanded into terms with up to 4 gluons as
\begin{equation}
D_N({k_1},k_2)=D_0(k_1) \,(2\pi)^4\,\delta^4({k_1}-k_2)+{\Sigma({k_1},k_2)}
\end{equation}
$D_0(k_1)$ is the inverse propagator, 
\begin{equation}
D_0(k_1) = 1+\frac{\chi_0(k_1)}{\omega(k_1)}, \qquad \omega(p)=\sqrt{\Bigl(\sum_\mu \sin^2{(p_\mu)}\Bigr)+\Bigl(\rho - 2r \sum_\mu \sin^2{(p_\mu/2)}\Bigr)^2}
\label{D0_omega}
\end{equation}
and
\begin{eqnarray}
\frac{\Sigma({k_1},k_2)}{\rho} &&= \underbrace{V_1^1({k_1},k_2)}_{\rm 1-gluon\,vertex} +
                                   \underbrace{V_1^2({k_1},k_2)+V_2^2({k_1},k_2)}_{\rm 2-gluon\, vertex} \nonumber \\
\nonumber \\
                               &&+ \underbrace{V_1^3({k_1},k_2)+V_2^3({k_1},k_2)+V_3^3({k_1},k_2)}_{\rm 3-gluon\, vertex} \nonumber \\
\nonumber \\
                               &&+ \underbrace{V_1^4({k_1},k_2)+V_2^4({k_1},k_2)+V_3^4({k_1},k_2)+V_4^4({k_1},k_2)}_{\rm 4-gluon\, vertex} 
                                 +{\cal O}(g_0^5)
\label{vertices}
\end{eqnarray}
where we have set $a=1$.
$V_1^i$,$V_2^i$, $V_3^i$ are given below and the reader can find the expression for $V_4^4$ in Appendix A.
\begin{eqnarray}
\hskip -1cm
V_1^i({k_1}&&,k_2) = \Brown{\frac{1}{\omega({k_1})+\omega(k_2)}}\Red{\Bigg[}X_i({k_1},k_2)-
                            \Brown{\frac{1}{\omega({k_1})\omega(k_2)}}\chi_0({k_1})\,X_i^{\dagger}({k_1},k_2)\,\chi_0(k_2)\Red{\Bigg]} 
\label{V1}\\
\hskip -.75cm 
\phantom{---} \nonumber \\
\hskip -.75cm 
\phantom{---} \nonumber \\
\hskip -.75cm
V_2^i({k_1}&&,k_2) =\int\frac{d^4k_3}{(2\pi)^4}\,\Brown{\frac{1}{\omega(k_1)+\omega(k_3)}\frac{1}{\omega(k_1)+\omega(k_2)}
\frac{1}{\omega(k_2)+\omega(k_3)}}\times \nonumber \\
\hskip -.75cm
&& \sum_{\{j>0,k>0\} \atop \{j+k=i\}} \Red{\Bigg[}-X_j(k_1,k_3)\,\chi_0^{\dagger}(k_3)\,X_k(k_3,k_2)\nonumber \\[-3.0ex]
\hskip -.75cm
&& \qquad\quad\quad-X_j(k_1,k_3)\,X_k^{\dagger}(k_3,k_2)\,\chi_0(k_2)\nonumber \\
\hskip -.75cm
&& \qquad\quad\quad-\chi_0(k_1)\,X_j^{\dagger}(k_1,k_3)\,X_k(k_3,k_2)\nonumber \\
\hskip -.75cm
&& \qquad\quad\quad+\Brown{\frac{\omega(k_1)+\omega(k_2)+\omega(k_3)}{\omega(k_1)\omega(k_2)\omega(k_3)}}
\chi_0(k_1)X_j^{\dagger}(k_1,k_3)\chi_0(k_3)X_k^{\dagger}(k_3,k_2)\chi_0(k_2)\Red{\Bigg]}
\label{V2}\\
\hskip -.75cm
\phantom{---} \nonumber \\
\hskip -.75cm 
\phantom{---} \nonumber \\
\hskip -.75cm
V_3^i({k_1}&&,k_2) =\int\int{d^4 k_3\over(2\pi)^4}{d^4k_4\over(2\pi)^4}
{1\over4}\Brown{\Biggl(\prod_{p \epsilon S_4}{1\over\omega(k_{p_1})+\omega(k_{p_2})}\Biggr)}\times\nonumber\\
\hskip -.75cm
&& \sum_{\{j>0,k>0,l>0\} \atop \{j+k+l=i\}} \Red{\Biggl[}
-\frac{1}{6}\Brown{\Biggl(\sum_{p \epsilon S_4} \omega(k_{p_1})\omega(k_{p_2})\omega(k_{p_3})\Biggr)} X_j(k_1,k_3)X_k^{\dagger}(k_3,k_4)X_l(k_4,k_2) \nonumber\\
\hskip -.75cm
&& + \Brown{\Bigl(\omega(k_1)+\omega(k_3)+\omega(k_4)+\omega(k_2)\Bigr)}\times\nonumber\\
\hskip -.75cm
&& \Blue{\Bigl[}\chi_0(k_1)X_j^{\dagger}(k_1,k_3)X_k(k_3,k_4)X_l^{\dagger}(k_4,k_2)\chi_0(k_2)+
\chi_0(k_1)X_j^{\dagger}(k_1,k_3)X_k(k_3,k_4)\chi_0^{\dagger}(k_4)X_l(k_4,k_2)\nonumber\\
\hskip -.75cm
&&+  \chi_0(k_1)X_j^{\dagger}(k_1,k_3)\chi_0(k_3)X_k^{\dagger}(k_3,k_4)X_l(k_4,k_2)+ 
X_j(k_1,k_3)\chi_0^{\dagger}(k_3)X_k(k_3,k_4)X_l^{\dagger}(k_4,k_2)\chi_0(k_2)\nonumber\\
\hskip -.75cm
&&+X_j(k_1,k_3)\chi_0^{\dagger}(k_3)X_k(k_3,k_4)\chi_0^{\dagger}(k_4)X_l(k_4,k_2)+
X_j(k_1,k_3)X_k^{\dagger}(k_3,k_4)\chi_0(k_4)X_l^{\dagger}(k_4,k_2)\chi_0(k_2)\Blue{\Bigr]}\nonumber\\
\hskip -.75cm
&&-\Brown{\Biggl(\sum_{p \epsilon S_4} {\omega(k_{p_1})\omega(k_{p_2}) \Bigl(\omega(k_{p_1})/2+\omega(k_{p_3})/3 \Bigr)
\over\omega(k_1)\omega(k_3)\omega(k_4)\omega(k_2)}\Biggr)}\times \nonumber\\
\hskip -.75cm
&&{\chi_0(k_1)X_j^{\dagger}(k_1,k_3)\chi_0(k_3)X_k^{\dagger}(k_3,k_4)\chi_0(k_4)X_l^{\dagger}(k_4,k_2)\chi_0(k_2)}\Red{\Biggr]}
\label{V3}
\end{eqnarray}
($S_4$: permutation group of the numbers $\{1,2,3,4\}$)

The use of the background field technique implies that instead of the generic gluonic fields 
(appearing in $X_i$'s, Eq.~(\ref{X's})), one must consider all possible combinations of background 
(\Red{A}) and quantum (\Blue{Q}) fields which originate in the links of Eqs.~(\ref{derivative}), (\ref{link}). Hence,
\begin{eqnarray}
&&X_1(p',p)= X_1^{\Blue{Q}}(p',p)+X_1^{\Red{A}}(p',p) \nonumber \\
&&X_2(p',p)= X_2^{\Blue{Q}\Blue{Q}}(p',p)+X_2^{\Blue{Q}\Red{A}}(p',p)+X_2^{\Red{A}\Red{A}}(p',p) \nonumber\\
&&X_3(p',p)= X_3^{\Blue{Q}\Blue{Q}\Blue{Q}}(p',p)+X_3^{\Blue{Q}\Blue{Q}\Red{A}}(p',p)+
X_3^{\Blue{Q}\Red{A}\Red{A}}(p',p)+X_3^{\Red{A}\Red{A}\Red{A}}(p',p) \nonumber\\
&&X_4(p',p)= X_4^{\Blue{Q}\Blue{Q}\Blue{Q}\Blue{Q}}(p',p)+X_4^{\Blue{Q}\Blue{Q}\Blue{Q}\Red{A}}(p',p)+
X_4^{\Blue{Q}\Blue{Q}\Red{A}\Red{A}}(p',p)+X_4^{\Blue{Q}\Red{A}\Red{A}\Red{A}}(p',p)+X_4^{\Red{A}\Red{A}\Red{A}\Red{A}}(p',p)
\label{X_QA}
\end{eqnarray}
As an example, let us write the expression for $X_3^{QQA}$ 
\begin{eqnarray}
X_3^{QQA}(p',&&p)= \frac{g_0^2}{4}\int
\frac{d^4k_1d^4k_2d^4k_3}{(2\pi)^8}\delta(p'-p-k_1-k_2-k_3)\times \nonumber \\
&&\sum_{\mu}\Bigg[-a^2 V_{1,\mu}\Big(\frac{p'+p}{2}\Big)\Big(Q_{\mu}(k_1)Q_{\mu}(k_2)A_{\mu}(k_3)+A_{\mu}(k_3)Q_{\mu}(k_2)Q_{\mu}(k_1)\Big) \nonumber \\
&&\phantom{spa\Bigg[}+i a V_{2,\mu}\Big(\frac{p'+p}{2}\Big)\Big(Q_{\mu}(k_1)Q_{\mu}(k_2)A_{\mu}(k_3)-A_{\mu}(k_3)Q_{\mu}(k_2)Q_{\mu}(k_1)\Big) \Bigg]
\end{eqnarray}
Upon substituting the expression for $X_i$'s in the overlap vertices, the latter become extremely lengthy and complicated. 
For instance, the vertex with Q-Q-A-$\Psi$-$\overline{\Psi}$ consists of 9,784 terms,
while the vertex with Q-Q-A-A-$\Psi$-$\overline{\Psi}$ has 724,120 terms.


\section{RESULTS}

For the algebra involving lattice quantities, we make use of our symbolic manipulation package in Mathematica, 
with the inclusion of the additional overlap vertices. The first step to evaluate the diagrams is the contraction among vertices, 
a step performed automatically once the vertices and the `incidence matrix' of the diagram are specified. 
The outcome of contraction is a preliminary expression for the diagram under study; 
there follow simplifications of the color dependence, $\rm Dirac$ matrices and tensor structures.
We use symmetries of the theory (permutation symmetry and lattice rotational invariance), or any other additional symmetry that may appear
in particular diagrams, to keep the size of the expression down to a minimum. 
The external momentum $p$ appears in arguments of trigonometric functions and 
the extraction of $p$ dependence is divided into two parts: first we isolate terms that give single and double logarithms 
(a few thousand terms, expressible in terms of known, tabulated integrals),
and then for the convergent terms we employ naive Taylor expansion with respect to $p$ up to ${\cal O}(p^2)$.
This extraction makes explicit the functional dependence of each diagram on $p$; 
the coefficients of terms proportional to $p^2$ are integrals over the two internal momentum 4-vectors. 
The required numerical integrations are performed by optimized Fortran programs which are generated by our Mathematica `integrator' routine. 
Each integral is expressed as a sum over the discrete Brillouin zone of finite lattices, with varying size $L$,
and evaluated for different values of the overlap parameter $\rho$.
The average length of the expression for each diagram, after simplifications, is about 2-3 hundred thousand terms, so that diagrams must be split into 
parts (usually of 2000 terms) to be integrated. The numerical values of these parts must then 
be added together to avoid running into systematic errors or spurious divergences.
Finally, we extrapolate the results to $L\to\infty$; this procedure introduces an inherent systematic error,
which we can estimate quite accurately. Infrared divergent diagrams must be summed up before performing the extrapolation.

\smallskip
We denote the contribution of the $i^{\rm th}$ 1-loop Feynman diagram to $\nu^{(1)}(p)$ as $\nu_i^{(1)}(p)$;
similarly, contributions of 2-loop diagrams to $\nu^{(2)}(p)$ are indicated by $\nu_i^{(2)}(p)$.
The quantities $\nu_i^{(1)}(p)$, $\nu_i^{(2)}(p)$ depend on $N$, $N_f$, $\rho$ and $ap$ according to the following formulae ($\lambda_0=1$)
\begin{eqnarray}
\hskip -.2cm 
&&\widehat{ap}^2 \nu_i^{(1)}(p) = N_f \Biggl[k_{i}^{(0)} + a^2 p^2 \Bigl\{k_{i}^{(1)} + 
k_{i}^{(2)} {\ln a^2 p^2 \over (4 \pi)^2} \Bigr \} + {\cal O}((ap)^4) \Biggr]
\label{nu1Lattice}\\
\nonumber \\
&&\widehat{ap}^2 \nu_i^{(2)}(p) = N_f \Biggl[c_{i}^{(0)} + a^2 p^2 \left\{c_{i}^{(1)}+ c_{i}^{(2)}\, {\ln a^2 p^2 \over (4 \pi)^2} +
c_{i}^{(3)} \left({\ln a^2 p^2 \over (4 \pi)^2}\right)^2  +c_{i}^{(4)}{\displaystyle\sum_\mu p_\mu^4 \over (p^2)^2} \right \} + {\cal O}((ap)^4)\Biggr]
\label{nu2Lattice}
\end{eqnarray}
where $\widehat{p}^2 = 4 \sum_\mu \sin^2(p_\mu/2)$. 
The index $i$ runs over diagrams, and the coefficients $k_{i}^{(j)}$, $c_{i}^{(j)}$ depend on the overlap parameter $\rho$.
Moreover, $c_{i}^{(j)}= \Bigl[ c_{i}^{(j,-1)}/N + c_{i}^{(j,1)}N \Bigr]$. Comparison with continuum results and usage of Ward Identities 
requires

\begin{eqnarray}
&&\bullet\,\, \sum_i k_{i}^{(0)} =0,\,\qquad\,\sum_i c_{i}^{(0)} = 0 \qquad {\rm (gauge\,\, invariance)}\nonumber\\
&&\bullet\,\, \sum_i k_{i}^{(2)} = \frac{2}{3},\,\qquad\,\sum_i c_{i}^{(2)} = {1\over 16 \pi^2} (3 N - {1\over N})\nonumber\\
&&\bullet\,\, \sum_{i} c_{i}^{(4)}=0 \qquad\,\, {\rm (Lorentz\,\, invariance)}\nonumber\\
&&\bullet\,\, c_{15}^{(3)} = {1\over 3 N}  , \quad c_{16}^{(3)}  = {4\over 3} N  , \quad 
                      c_{17}^{(3)}  = -{5\over 3} N  , \quad c_{18}^{(3)}  = {N^2-1\over 3 N} \qquad\qquad\qquad\qquad\qquad
\end{eqnarray}
We have checked that all the above conditions are verified by our results.
Inserting these conditions in Eqs. (\ref{nu1Lattice}), (\ref{nu2Lattice}), 
the expressions for the fermionic contribution to $\nu^{(1)}(p)$, $\nu^{(2)}(p)$ 
(after addition of all diagrams) take the form
\begin{eqnarray}
&&\nu^{(1)}(p) =  \nu^{(1)}(p)\Big|_{N_f=0} +  N_f \Biggl[\sum_i k_{i}^{(1)} + 
\frac{2}{3} {\ln a^2 p^2 \over (4 \pi)^2} + {\cal O}((ap)^2) \Biggr]
\label{nu1Lattice_final}\\
\nonumber \\
&&\nu^{(2)}(p) = \nu^{(2)}(p)\Big|_{N_f=0} +  N_f \Biggl[\sum_i \Big({c_{i}^{(1,1)}\over N} + N c_{i}^{(1,1)} \Big) +
{1\over 16 \pi^2} (3 N - {1\over N})\, {\ln a^2 p^2 \over (4 \pi)^2} + {\cal O}((ap)^2)\Biggr]
\label{nu2Lattice_final}
\end{eqnarray}

In Table I we tabulate the total 1-loop contribution $k^{(1)}\equiv\sum_i k_{i}^{(1)}$ 
for 21 values of the overlap parameter ($0<\rho<2$). Each diagram was integrated for lattice size $L^4$, 
$L \leq 128$ and then extrapolated to $L \rightarrow\infty$. 
In all Tables and Figures, the errors accompanying our results are entirely due to this extrapolation.
The coefficients $k_{i}^{(1)}(\rho)$ do not depend on the number of colors $N$, nor on  
the choice of regularization for the pure gluonic part of the action (Symanzik, Iwasaki, etc.).
The numbers in Table I are in agreement with corresponding numbers from Ref.~\cite{APV}.

The 2-loop calculation of $\nu^{(2)}(p)$ was accomplished for the same 21 values of $\rho$ and for $L \leq 28$.
Table II presents the coefficients $c^{(1,-1)}\equiv\sum_i c_{i}^{(1,-1)}$, 
$c^{(1,1)}\equiv\sum_i c_{i}^{(1,1)}$, versus $\rho$.
Due to the extremely large size of the vertices involved, it is almost impossible to extend the results to larger $L$.
Typically, the integration of 2000 terms is completed in  $\sim$ 7 days on 1 CPU;
the present calculation comprises approximately 3500$\times$2000 terms. 
Thus, if only a single CPU were available, our work would have required $\sim$50 years. 
In certain cases with large systematic errors we extended the integration up to $L=46$.
In general, the overlap action leads to coefficients which are very small for most values of $\rho$. 
As a consequence, systematic errors, which are by and large rather small, tend to 
be significant fractions of the signal for $\rho > 1.4$.

From Eqs.~(\ref{Zg}), (\ref{Zg_expanded})-(\ref{omega1_1}), (\ref{nu2_1}) 
we find the following expression for $l_1$ in terms of $N,\,N_f,$ $k_{i}^{(1)}$, $c_{i}^{(1,-1)}$, $c_{i}^{(1,1)}$
\begin{eqnarray}
l_1 =&& -\frac{3}{128\,N^2} +0.018127763034 - 0.007910118514\, N^2 \nonumber \\
&&+N_f \Bigg[\frac{1}{(16\pi^2)^2 N} \Big({55\over 12} -4\zeta(3) \Big) -\frac{N}{(16\pi^2)^2}{481\over 36}
-\frac{N}{8\pi^2} k^{(1)} - \Big({c^{(1,-1)}\over N} + N c^{(1,1)}  \Big)\Bigg]
\label{l1}
\end{eqnarray}
We can write the final form of the 3-loop coefficient $b_2^L$ for the $\beta$-function (Eq.~(\ref{b2lrel})),
including gluonic as well as fermionic contributions, 
using Eqs. (\ref{univ_coeff}), (\ref{b2MS}), (\ref{l0}) and (\ref{l1})
\begin{eqnarray}
b_2^L =&& -{11 \over 2048\pi^2\,N} + 0.000364106020\, N -0.000092990690\, N^3 \nonumber \\
\nonumber \\
+&&N_f \Bigg[\frac{(4\pi^2-1)^2}{4 (16\pi^2)^3\,N^2} - 0.000046883436 -  000013419574\,N^2  \nonumber \\
\nonumber \\
&&\quad\quad +\frac{N_f}{(16\pi^2)^3}\Bigg( -{23\over 9\,N}+{8\zeta(3)\over 3\,N}+{37\,N\over 6}\Bigg)\nonumber \\
\nonumber \\
&&\quad\quad -\frac{(11\,N-2\,N_f)}{48\pi^2}\Big({c^{(1,-1)}\over N} + c^{(1,1)}\,N \Big) 
+ \frac{(4\,N^3+N_f-3\,N^2\,N_f)}{(16\pi^2)^2\,N}k^{(1)} \Bigg]
\label{b2_final}
\end{eqnarray}

A large variety of possible numerical checks has been performed, as mentioned above:
{\bf a.} The total contribution to the gluon mass adds to zero, as expected. 
{\bf b.} The coefficients of the non-Lorentz invariant terms cancel.
{\bf c.} The terms with double logarithms correspond to the continuum counterparts. 
This has been checked diagram by diagram.
{\bf d.} Terms with single logarithms add up to their expected value, which is independent of $\rho$ 
(although the expressions per diagram are $\rho$-dependent).

In Fig. 3 we plot the 1-loop coefficient $k^{(1)}$  with respect to $\rho$. 
Note that the errors are too small to be visible at this scale. The 2-loop coefficients 
$c^{(1,-1)}$ and $c^{(1,1)}$ are plotted in Figs. 4-5, respectively,
for different values of the overlap parameter. 
The extrapolation errors are visible for $\rho\leq 0.4$ and $\rho\geq 1.7$.
Substituting $k^{(1)}$, $c^{(1,-1)}$ and $c^{(1,1)}$ into 
Eq.~(\ref{b2_final}), we find the numerical results for the 3-loop contribution, $b_2^L$,
of the $\beta$-function. These are plotted in Fig. 6, choosing $N=3$ and $N_f=0,\,2,\,3$.


\vskip 1cm
\centerline{\psfig{file=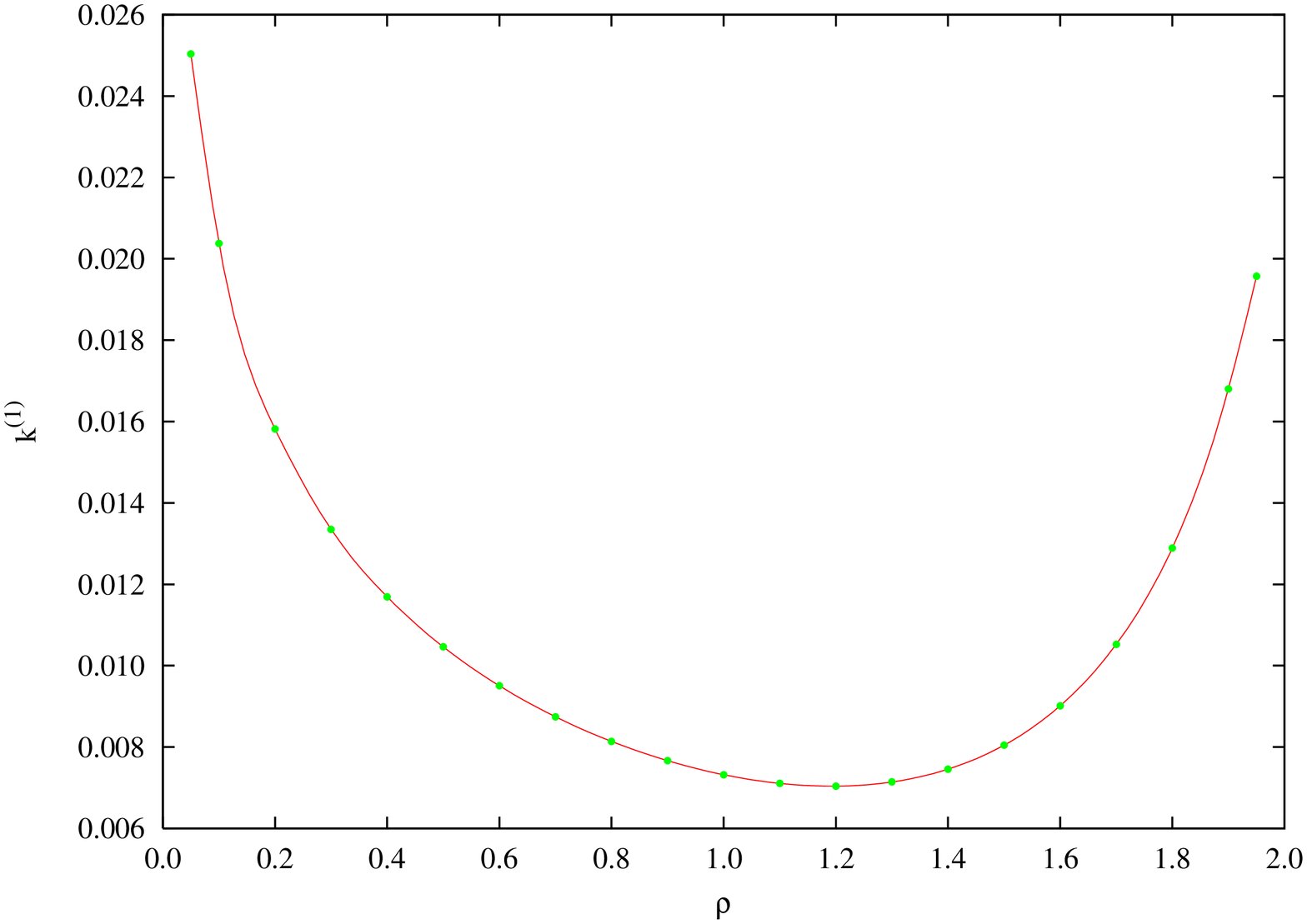,scale=.55,angle=-90}}
\vskip 2mm
\hskip -5mm
\centerline{\footnotesize Fig. 3: Plot of the total 1-loop coefficient $\displaystyle k^{(1)}\equiv\sum_i k_{i}^{(1)}$ versus the overlap parameter $\rho$.}
\vskip 1cm

\vskip .5cm
\centerline{\psfig{file=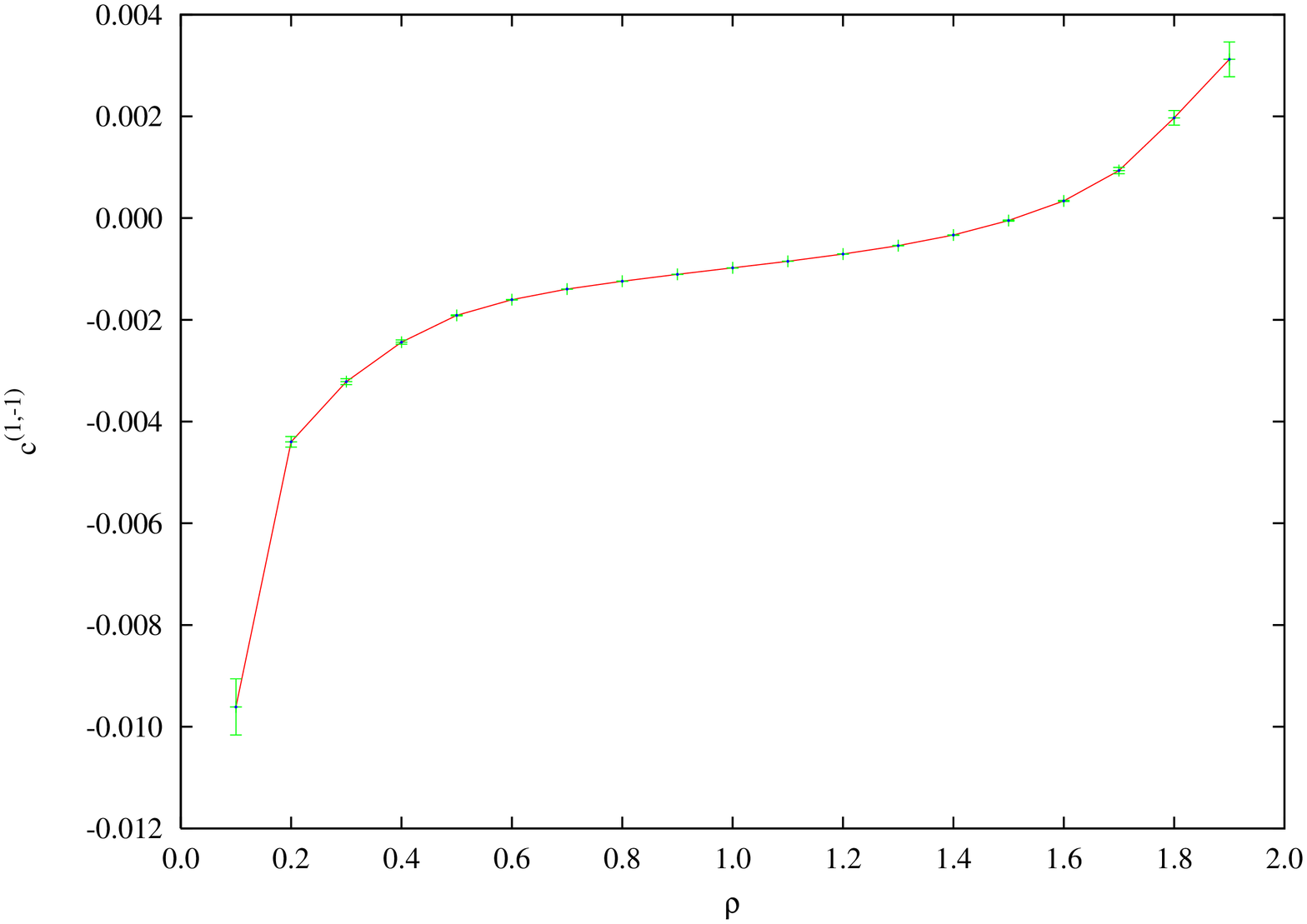,scale=.55,angle=-90}}
\vskip 2mm
\centerline{\footnotesize Fig. 4: Plot of the total 2-loop coefficient $\displaystyle c^{(1,-1)}\equiv\sum_i c_i^{(1,-1)}$ versus $\rho$.}
\vskip .5cm
\centerline{\psfig{file=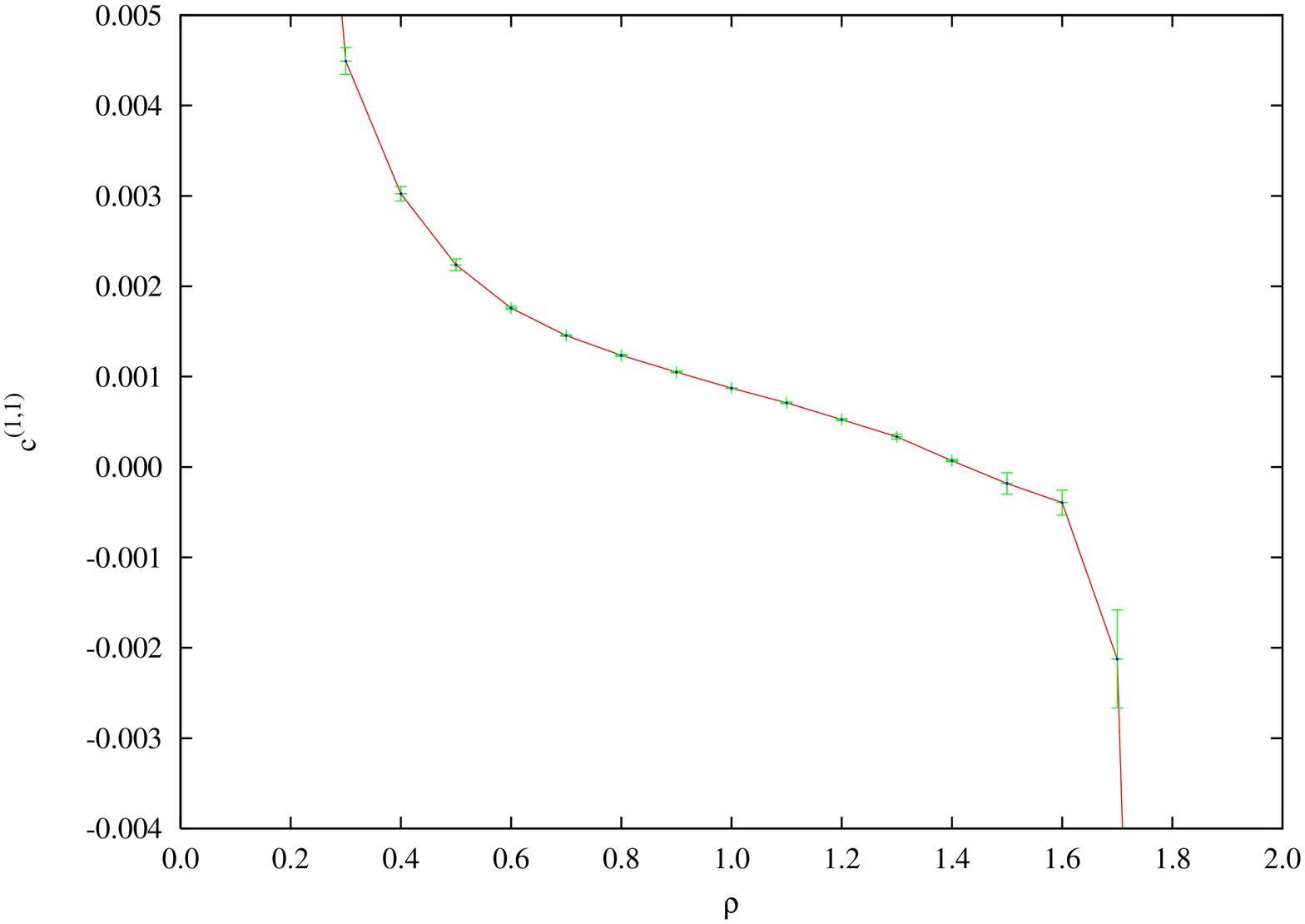,scale=.55,angle=-90}}
\vskip 2mm
\centerline{\footnotesize Fig. 5: Plot of the total 2-loop coefficient $\displaystyle c^{(1,1)}\equiv\sum_i c_i^{(1,1)}$ versus $\rho$.}
\vskip .5cm
\vskip .5cm
\centerline{\psfig{file=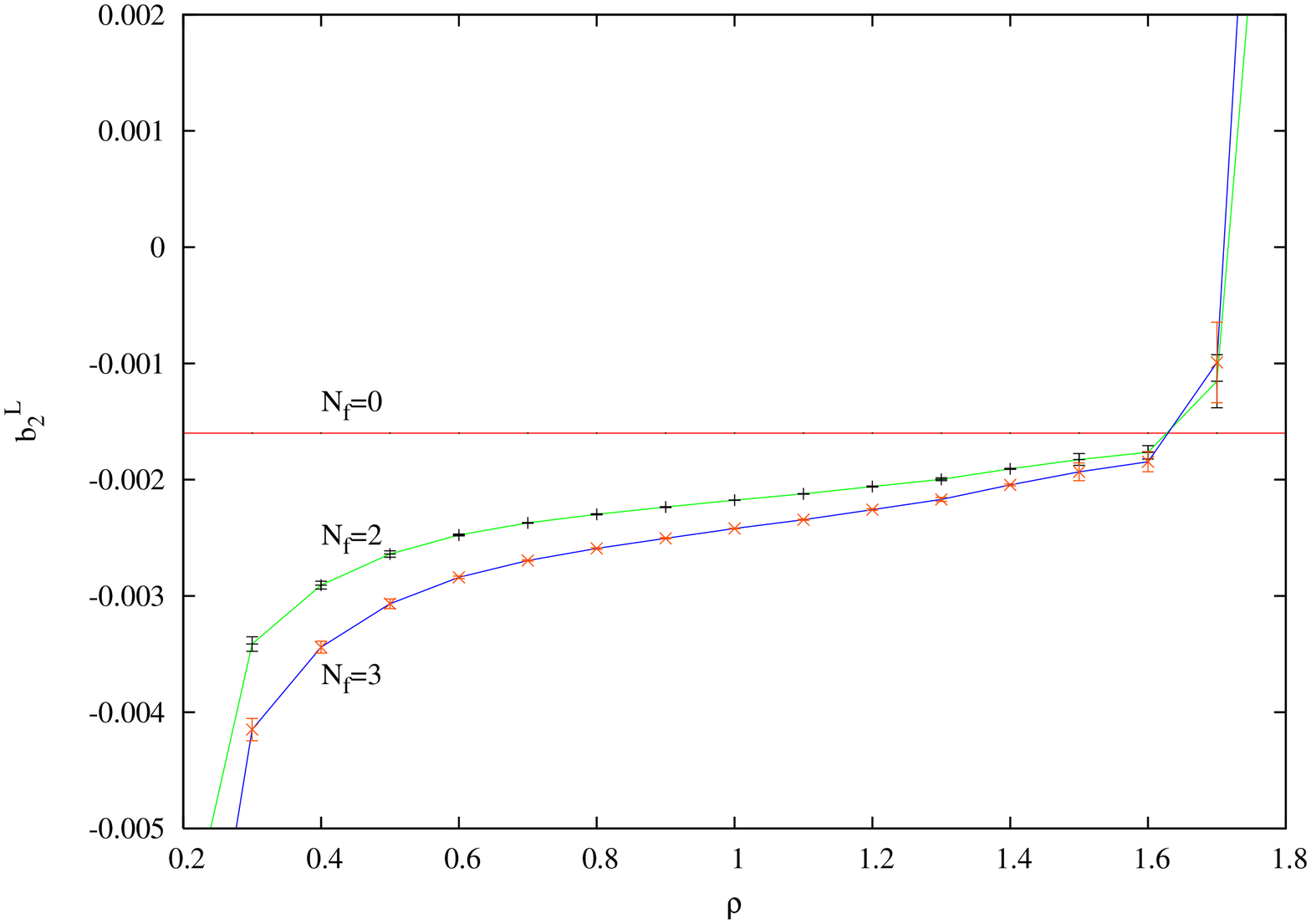,scale=.55,angle=-90}}
\vskip 2mm
\noindent
       {\footnotesize Fig. 6: The 3-loop coefficient $b_2^L$ (Eq. (\ref{b2_final})), plotted against $\rho$, for $N=3$ and $N_f=0$ (horizontal red line),
$N_f=2$ (green line) and $N_f=3$ (blue line).}

\section{SUMMARY AND CONCLUSIONS}

We have calculated the 2-loop coefficient of the coupling renormalization function $Z_g$, 
for the Yang-Mills theory with gauge group $SU(N)$ and $N_f$ species of overlap fermions.
We used the background field method to simplify the computation; in
this method there is no need of evaluating any 3-point functions. 
This is the first 2-loop calculation using overlap fermions with
external momenta, and it proved to be extremely demanding in human and
CPU time; this is due to the fact that
we had to manipulate very large expressions (millions of terms) in intermediate stages. 

We used our numerical results of $Z_g$ to determine the 3-loop
coefficient $b_2^L$ of the bare lattice $\beta$-function; the latter
dictates the asymptotic dependence between the bare coupling constant $g_0$
and the lattice spacing $a$, required to maintain the
renormalized coupling at a given scale fixed. Knowledge of $b_2^L$ provides
the correction term to the standard asymptotic scaling relation
between $a$ and $g_0$, via Eq. (\ref{asympre}).

The dependence of $Z_g$ and $b_2^L$ on $N$ and $N_f$ is shown explicitly
in our expressions. On the other hand, dependence on the overlap parameter
$\rho$ cannot certainly be given in closed form; instead, we present
our results for a large set of values of $\rho$ in its allowed range.

The 3-loop correction is seen to be rather small: This indicates that
the perturbative series is very well behaved in this case, despite the
fact that it is only asymptotic in nature.
Furthermore, around the values of $\rho$ which are most often used in
simulations ($1\leq\rho\leq 1.6$), fermions bring about only slight
corrections to the 3-loop $\beta$-function, even compared to pure gluonic contributions, 
as can be seen from Fig. 6.

The only source of numerical error in our results has its origin in an
extrapolation to infinite lattice size. Compatibly with the severe CPU
constraints, numerical 2-loop integration had to
be performed on lattices typically as large as $28^4$, or even up to
$46^4$ in cases where an improved extrapolation was called for. 
An intermediate range for $\rho$ ($0.6\leq\rho\leq 1.3$) showed the
most stable extrapolation error, and this may be a sign of their
suitability for numerical simulations.

As a by product of the present work, we have produced the lengthy
expressions corresponding to {\it all} overlap vertices which can arise in a
2-loop computation, and presented them in a rather compact
form. Further computations of similar complexity, for example the
2-loop renormalization of operators in the overlap action (such as fermion
currents), only
require the vertices which we have presented here.

\bigskip\noindent
{\bf Acknowledgements: } This work is supported in part by the
Research Promotion Foundation of Cyprus (Proposal Nr: $\rm ENI\Sigma X$/0505/45).

\textheight=25cm
\newpage
\begin{table}[tbp]
\begin{minipage}{5cm}
\hfill
\end{minipage}
\begin{minipage}{7cm}
\label{tab1}
\begin{tabular}{|c|l|}
{\phantom{space}$\rho$\phantom{space}}&$ \hskip -.5cm {k^{(1)}}$\\
\tableline \hline
0.1  &\hskip -1.5cm $$0.020377(7) \phantom{space}\\
0.2  &\hskip -1.5cm $$0.01581702(2) \phantom{space}\\
0.3  &\hskip -1.5cm $$0.0133504717(2) \phantom{space}  \\
0.4  &\hskip -1.5cm $$0.0116910952(1) \phantom{space}   \\
0.5  &\hskip -1.5cm $$0.0104621922(2) \phantom{space}  \\
0.6  &\hskip -1.5cm $$0.0095058191(2) \phantom{space}  \\
0.7  &\hskip -1.5cm $$0.00874441051(7) \phantom{space}   \\
0.8  &\hskip -1.5cm $$0.00813753230(4) \phantom{space}   \\
0.9  &\hskip -1.5cm $$0.00766516396(3) \phantom{space}   \\
1.0  &\hskip -1.5cm $$0.00732057894(3)\phantom{space}  \\
1.1  &\hskip -1.5cm $$0.00710750173(2) \phantom{space}   \\
1.2  &\hskip -1.5cm $$0.00703970232(7) \phantom{space}   \\
1.3  &\hskip -1.5cm $$0.0071425543(2) \phantom{space}  \\
1.4  &\hskip -1.5cm $$0.0074569183(2) \phantom{space}  \\
1.5  &\hskip -1.5cm $$0.0080467046(1) \phantom{space}   \\
1.6  &\hskip -1.5cm $$0.0090134204(1) \phantom{space}  \\
1.7  &\hskip -1.5cm $$0.010526080(2)  \phantom{space} \\
1.8  &\hskip -1.5cm $$0.0128914(2)  \phantom{space}   \\
1.9  &\hskip -1.5cm $$0.01680(8)  \phantom{space} 
\end{tabular}
\end{minipage}
\vskip .25cm
\caption{Coefficient $\displaystyle {k^{(1)}}\equiv\sum_i k_{i}^{(1)}$ for different values of the overlap parameter $\rho$.}
\vskip .5cm
\begin{minipage}{3cm}
\hfill
\end{minipage}
\begin{minipage}{10cm}
\label{tab2}
\begin{tabular}{|c|l|l|}
{\phantom{space}$\rho$\phantom{space}}&$ \hskip -1.15cm c^{(1,-1)}$ &$ \hskip -1.15cm c^{(1,1)}$ \\
\tableline \hline
0.1  &\hskip -1.5cm -$0.$0096(6) \phantom{space}&\hskip -1.5cm \phantom{-}$$0.124(3) \phantom{space}\\
0.2  &\hskip -1.5cm -$0.$0044(1) \phantom{space}&\hskip -1.5cm \phantom{-}$$0.0118(5) \phantom{space}\\
0.3  &\hskip -1.5cm -$0.$00321(6) \phantom{space} &\hskip -1.5cm \phantom{-}$$0.0045(1) \phantom{space}  \\
0.4  &\hskip -1.5cm -$0.$00244(4) \phantom{space} &\hskip -1.5cm \phantom{-}$$0.0030(1) \phantom{space}  \\
0.5  &\hskip -1.5cm -$0.$00191(1) \phantom{space}  &\hskip -1.5cm \phantom{-}$$0.0022(6) \phantom{space} \\
0.6  &\hskip -1.5cm -$0.$001606(6) \phantom{space} &\hskip -1.5cm \phantom{-}$$0.00176(2) \phantom{space}  \\
0.7  &\hskip -1.5cm -$0.$001397(3) \phantom{space} &\hskip -1.5cm \phantom{-}$$0.00145(1) \phantom{space}  \\
0.8  &\hskip -1.5cm -$0.$001241(1) \phantom{space} &\hskip -1.5cm \phantom{-}$$0.00124(1) \phantom{space}  \\
0.9  &\hskip -1.5cm -$0.$001107(1) \phantom{space} &\hskip -1.5cm \phantom{-}$$0.001051(9) \phantom{space}  \\
1.0  &\hskip -1.5cm -$0.$000979(1)\phantom{space} &\hskip -1.5cm \phantom{-}$$0.000872(3)\phantom{space}  \\
1.1  &\hskip -1.5cm -$0.$000849(2) \phantom{space} &\hskip -1.5cm \phantom{-}$$0.000710(8) \phantom{space} \\
1.2  &\hskip -1.5cm -$0.$000706(3) \phantom{space} &\hskip -1.5cm \phantom{-}$$0.00052(1) \phantom{space} \\
1.3  &\hskip -1.5cm -$0.$000543(4) \phantom{space} &\hskip -1.5cm \phantom{-}$$0.00033(3) \phantom{space} \\
1.4  &\hskip -1.5cm -$0.$000335(7) \phantom{space} &\hskip -1.5cm \phantom{-}$$0.00007(1)\phantom{space}  \\
1.5  &\hskip -1.5cm -$0.$00005(1)\phantom{space} &\hskip -1.5cm -$0.$0002(1) \phantom{space}  \\
1.6  &\hskip -1.5cm \phantom{-}$$0.00034(1) \phantom{space} &\hskip -1.5cm -$0.$0004(1)\phantom{space}  \\
1.7  &\hskip -1.5cm \phantom{-}$$0.00093(6)  \phantom{space} &\hskip -1.5cm -$0.$0021(5) \phantom{space}\\
1.8  &\hskip -1.5cm \phantom{-}$$0.0020(1)  \phantom{space}&\hskip -1.5cm -$0.$02(3) \phantom{space} 
\end{tabular}
\end{minipage}
\vskip .25cm
\caption{Numerical results for the 2-loop coefficients $\displaystyle c^{(1,-1)}\equiv\sum_i c_{i}^{(1,-1)}$ and $\displaystyle c^{(1,1)}\equiv\sum_i c_{i}^{(1,1)}$}
\end{table}

\newpage

\appendix
\section{Construction of vertices}

Here we will explain the basic idea of expanding the expressions $\displaystyle 1 / \sqrt{X^\dagger X}$, 
appearing in the overlap-Dirac operator, in powers of $g_0$, 
using a procedure introduced by Y. Kikukawa and A. Yamada~\cite{KY}.
We also provide the expression for $V_4^4({k_1},k_2)$ further below.

In an integral representation, the combination $\displaystyle \frac{1}{\sqrt{X^\dagger X}}$ can be written as
\begin{equation}
\frac{1}{\sqrt{X^\dagger X}} = \int^\infty_{-\infty} \frac{dt}{\pi}   \frac{1}{t^2 + X^\dagger X}
\label{XXdag}
\end{equation}
In fact, Eq.~(\ref{XXdag}) is valid for any operator $X$ provided that $X^\dagger X$ has no vanishing eigenvalues.
We begin the desired expansion of the overlap-Dirac operator in powers of $g_0$, by setting
\begin{equation}
X^\dagger X = \underbrace{X_0^\dagger X_0}_{{\cal O}(g_0^0)} +Z
\label{XXdag_expansion}
\end{equation}
The first term of Eq.~(\ref{XXdag_expansion}) corresponds to the inverse fermionic propagator, while $Z$ leads to the vertices;
for our 2-loop calculation we need to write $Z$ up to ${\cal O}(g_0^4)$
\begin{eqnarray}
Z =&& \underbrace{(X_0^\dagger X_1 + X_1^\dagger X_0 )}_{{\cal O}(g_0^1)} + 
\underbrace{(X_0^\dagger X_2 + X_1^\dagger X_1 + X_2^\dagger X_0)}_{{\cal O}(g_0^2)} + 
\underbrace{(X_0^\dagger X_3 + X_1^\dagger X_2 + X_2^\dagger X_1 + X_3^\dagger X_0)}_{{\cal O}(g_0^3)} \nonumber \\
+&&\underbrace{(X_0^\dagger X_4 + X_1^\dagger X_3 + X_2^\dagger X_2 + X_3^\dagger X_1 +  X_4^\dagger X_0)}_{{\cal O}(g_0^4)} + {\cal O}(g_0^5)
\end{eqnarray}
Using the above equations, we write the denominator on the r.h.s. of Eq.~(\ref{XXdag}) as
\begin{equation}
 \frac{1}{t^2 + X^\dagger X} =  \frac{1}{t^2 + X_0^\dagger X_0} 
\Bigl(1-Z\frac{1}{t^2 + X_0^\dagger X_0} + Z\frac{1}{t^2 + X_0^\dagger X_0}Z\frac{1}{t^2 + X_0^\dagger X_0} + ... \Bigr)
\label{tXXdag}
\end{equation}
From this point forward it is easier to work in momentum space since, taking the Fourier transform, the denominator becomes diagonal 
\begin{equation}
F.T.[\frac{1}{t^2 + X_0^\dagger X_0} ] = \frac{1}{t^2 + \omega^2(p)} 
\end{equation}
($\omega^2(p)$ defined in Eq.~(\ref{D0_omega})). Combining Eqs.~(\ref{XXdag}) and~(\ref{tXXdag}) 
we derive the Taylor expansion of $\displaystyle \frac{1}{\sqrt{X^\dagger X}}$
in momentum space
\begin{eqnarray}
\frac{1}{\sqrt{X^\dagger X}}_{F.T.}(p',p) =&&\int^\infty_{-\infty}\frac{dt}{\pi}\frac{2\pi\delta(p'-p)}{t^2 + \omega^2(p')}
-\int^\infty_{-\infty}\frac{dt}{\pi}\frac{1}{t^2 + \omega^2(p')}Z(p',p)\frac{1}{t^2 + \omega^2(p)} \nonumber \\
\vspace{.4cm}\nonumber \\
+&&\int^\infty_{-\infty}\frac{dt}{\pi}\int^\infty_{-\infty}\frac{dk}{(2\pi)^4}\frac{1}{t^2 + \omega^2(p')}Z(p',k)
\frac{1}{t^2 + \omega^2(k)}Z(k,p)\frac{1}{t^2 + \omega^2(p)} + ...
\label{SqrtXXdag_momentum}
\end{eqnarray}

\newpage
The integral over $t$ can now be performed by closing the contour around the upper complex $t$-plane. 
Considering, as an example the third term on the r.h.s. of Eq.~(\ref{SqrtXXdag_momentum}), the result is 
$$\int^\infty_{-\infty}\frac{dt}{\pi}\frac{1}{t^2 {+}\omega^2(p')}\,\frac{1}{t^2 {+}\omega^2(k)}\,\frac{1}{t^2 {+}\omega^2(p)} = \nonumber
\frac{\omega(p'){+}\omega(k){+}\omega(p)}{\omega(p')\omega(k)\omega(p) \nonumber
\left[\omega(p'){+}\omega(k)\right]\left[\omega(k){+}\omega(p)\right]\left[\omega(p){+}\omega(p')\right]}$$
Similarly we integrate all terms of Eq.~(\ref{SqrtXXdag_momentum}) over $t$ and this leads to Eqs.~(\ref{D0_omega})-(\ref{V3}) 
and to the expression for $V_4^4({k_1},k_2)$ which is presented below
{\small{
\begin{eqnarray}
V_4^4&&({k_1},k_2) =\int\int\int{d^4 k_3\over(2\pi)^4}{d^4k_4\over(2\pi)^4}{d^4 k_5\over(2\pi)^4}
{1\over12}\Brown{\Biggl(\prod_{p \epsilon S_5}{1\over\omega(k_{p_1})+\omega(k_{p_2})}\Biggr)}\times\nonumber\\
&& \Red{\Biggl[} \Brown{\Biggl(\sum_{p \epsilon S_5} \omega(k_{p_1})\omega(k_{p_2})\omega(k_{p_3})\omega(k_{p_4})
\Bigl( \omega(k_{p_1})/6 +\omega(k_{p_5})/30 \Bigr) \Biggr)}\times \nonumber\\
&&\Blue{\Bigl[}(X_1(k_1,k_3)X_1^\dagger(k_3,k_4)X_1(k_4,k_5)X_1^\dagger(k_5,k_2)\chi_0(k_2)\nonumber\\
&& +X_1(k_1,k_3)X_1^\dagger(k_3,k_4)X_1(k_4,k_5)\chi_0^\dagger(k_5)X_1(k_5,k_2) \nonumber\\
&& +X_1(k_1,k_3)X_1^\dagger(k_3,k_4)\chi_0(k_4)X_1^\dagger(k_4,k_5)X_1(k_5,k_2)\nonumber\\
&& +X_1(k_1,k_3)\chi_0^\dagger(k_3)X_1(k_3,k_4)X_1^\dagger(k_4,k_5)X_1(k_5,k_2) \nonumber\\
&& +\chi_0(k_1)X_1^\dagger(k_1,k_3)X_1(k_3,k_4)X_1^\dagger(k_4,k_5)X_1(k_5,k_2)\Blue{\Bigr]}\nonumber\\
&&-{1\over6}\Brown{\Biggl(\sum_{p \epsilon S_5} \omega(k_{p_1})\omega(k_{p_2}) \Bigl(\omega(k_{p_1})+\omega(k_{p_3}) \Bigr)\Biggr)}\times  \nonumber\\
&&\Blue{\Bigl[}X_1(k_1,k_3)\chi_0^\dagger(k_3)X_1(k_3,k_4)X_1^\dagger(k_4,k_5)\chi_0(k_5)X_1^\dagger(k_5,k_2)\chi_0(k_2)  \nonumber\\
&&+ X_1(k_1,k_3)\chi_0^\dagger(k_3)X_1(k_3,k_4)\chi_0^\dagger(k_4)X_1(k_4,k_5)\chi_0^\dagger(k_5)X_1(k_5,k_2) \nonumber\\
&&+ X_1(k_1,k_3)\chi_0^\dagger(k_3)X_1(k_3,k_4)\chi_0^\dagger(k_4)X_1(k_4,k_5)X_1^\dagger(k_5,k_2)\chi_0(k_2)\nonumber\\
&&+X_1(k_1,k_3)X_1^\dagger(k_3,k_4)\chi_0(k_4)X_1^\dagger(k_4,k_5)\chi_0(k_5)X_1^\dagger(k_5,k_2)\chi_0(k_2) \nonumber\\
&&+ \chi_0(k_1)X_1^\dagger(k_1,k_3)\chi_0(k_3)X_1^\dagger(k_3,k_4)\chi_0(k_4)X_1^\dagger(k_4,k_5)X_1(k_5,k_2)\nonumber\\
&&+\chi_0(k_1)X_1^\dagger(k_1,k_3)\chi_0(k_3)X_1^\dagger(k_3,k_4)X_1(k_4,k_5)\chi_0^\dagger(k_5)X_1(k_5,k_2) \nonumber\\
&&+ \chi_0(k_1)X_1^\dagger(k_1,k_3)\chi_0(k_3)X_1^\dagger(k_3,k_4)X_1(k_4,k_5)X_1^\dagger(k_5,k_2)\chi_0(k_2)\nonumber\\
&&+ \chi_0(k_1)X_1^\dagger(k_1,k_3)X_1(k_3,k_4)\chi_0^\dagger(k_4)X_1(k_4,k_5)\chi_0^\dagger(k_5)X_1(k_5,k_2)\nonumber\\
&& + \chi_0(k_1)X_1^\dagger(k_1,k_3)X_1(k_3,k_4)\chi_0^\dagger(k_4)X_1(k_4,k_5)X_1^\dagger(k_5,k_2)\chi_0(k_2)\nonumber\\
&&+\chi_0(k_1)X_1^\dagger(k_1,k_3)X_1(k_3,k_4)X_1^\dagger(k_4,k_5)\chi_0(k_5)X_1^\dagger(k_5,k_2)\chi_0(k_2)\Blue{\Bigr]}\nonumber\\
&&+\Brown{\Biggl(\sum_{p \epsilon S_5} {\omega^2(k_{p_1})\omega(k_{p_2})\omega(k_{p_3}) \over \omega(k_1)\omega(k_2)\omega(k_3)\omega(k_4)\omega(k_5)}\times}\nonumber\\
&&\Brown{\Bigl(\omega(k_{p_2}) [\omega(k_{p_1})/2+\omega(k_{p_3})/6]+ \omega(k_{p_4})[\omega(k_{p_1})/3 + \omega(k_{p_2})+\omega(k_{p_5})/3 ]\Bigr)\Biggr)}\times\nonumber\\
&& \chi_0(k_1)X_1^\dagger(k_1,k_3)\chi_0(k_3)X_1^\dagger(k_3,k_4)\chi_0(k_4)X_1^\dagger(k_4,k_5)\chi_0(k_5)X_1^\dagger(k_5,k_2)\chi_0(k_2)\Red{\Biggr]} 
\label{V4}
\end{eqnarray}
}}
($S_5$: permutation group of the numbers 1 through 5)

\textheight=25cm
\newpage

\end{document}